\documentclass[reprint,onecolumn,notitlepage,preprintnumbers,amsmath, aps,pre,amssymb]{revtex4-1}

\usepackage{comment}
\usepackage[caption=false]{subfig}
\usepackage{graphicx}

\usepackage[ruled,vlined,commentsnumbered]{algorithm2e}
\usepackage{adjustbox}
\usepackage{paralist}
\usepackage{lipsum}
\usepackage{bbm}
\usepackage{times}

\newcommand{\tg}[1]{}

\usepackage{expl3}
\ExplSyntaxOn
\newcommand\latinabbrev[1]{
  \peek_meaning:NTF . {
    #1\@}%
  { \peek_catcode:NTF a {
      #1.\@ }%
    {#1.\@}}}
\ExplSyntaxOff
\newcommand\etal{\latinabbrev{et al}}

\newtheorem{definition}{Definition}
\newcommand{\ignore}[1]{}

\newcommand\eg{\emph{e.g.}}
\newcommand\ie{\emph{i.e.}}

\newcommand\bcmdtab{\noindent\bgroup\tabcolsep=0pt%
  \begin{tabular}{@{}p{10pc}@{}p{20pc}@{}}}
\newcommand\ecmdtab{\end{tabular}\egroup}

 \newcommand{\remove}[1]{}
\usepackage{color}

\newcommand{\nop}[1]{}

\usepackage{color}

\usepackage[nolist]{acronym}

\begin{document}

\label{firstpage}

\title{Discriminative Predicate Path Mining for Fact Checking in Knowledge Graphs}

\author{Baoxu Shi}
 \email{bshi@nd.edu}
\author{Tim Weninger}
 \email{tweninge@nd.edu}
\affiliation{%
$^{\ast\dag}$Department of Computer Science and Engineering,
University of Notre Dame, Notre Dame, Indiana, USA
}%
\date{\today}

\begin{abstract}
Traditional fact checking by experts and analysts cannot keep pace with the volume of newly created information. It is important and necessary, therefore, to enhance our ability to computationally determine whether some statement of fact is true or false. We view this problem as a link-prediction task in a knowledge graph, and present a \emph{discriminative path}-based method for fact checking in knowledge graphs that incorporates connectivity, type information, and predicate interactions. Given a statement $\mathcal{S}$ of the form (\textsf{subject}, \textsf{predicate}, \textsf{object}), for example, (\textsf{Chicago}, \textsf{capitalOf}, \textsf{Illinois}), our approach mines discriminative paths that alternatively define the generalized statement (\textsf{U.S. city}, \textsf{predicate}, \textsf{U.S. state}) and uses the mined rules to evaluate the veracity of statement $\mathcal{S}$. We evaluate our approach by examining thousands of claims related to history, geography, biology, and politics using a public, million node knowledge graph extracted from Wikipedia and PubMedDB. Not only does our approach significantly outperform related models, we also find that the discriminative predicate path model is easily interpretable and provides sensible reasons for the final determination. 
\end{abstract}

\maketitle

\section{Introduction} \label{sec:introduction}

\begin{adjustbox}{minipage=0.82\linewidth,margin=0pt \smallskipamount,center}
        \begin{raggedright}
        \textit{\small{If a Lie be believ'd only for an Hour, it has done its Work, and there is no farther occasion for it. Falsehood flies, and the Truth comes limping after it.}}
        \end{raggedright}
        \begin{flushright}
        \vspace{-4pt}
        \small{-- Jonathan Swift (1710)~\cite{Swift1710}}
        \end{flushright}
        \vspace{2pt}
\end{adjustbox}

Misinformation in media and communication creates a situation in which opposing assertions of fact compete for attention. This problem is exacerbated in modern, digital society, where people increasingly rely on the aggregate ratings from their social circles for news and information. Although much of the information presented on the Web is a good resource, its accuracy certainly cannot be guaranteed. In order to avoid being fooled by false assertions, it is necessary to separate fact from fiction and to assess the credibility of an information source.

\vspace{5pt}\noindent{\textbf{Knowledge Graphs}.} We represent a \textit{statement of fact} in the form of (\textsf{subject}, \textsf{predicate}, \textsf{object}) triples, where the \textsf{subject} and the \textsf{object} are entities that have some relationship between them as indicated by the \textsf{predicate}. For example, the \textit{``Springfield is the capital of Illinois''} assertion is represented by the triple (\textsf{Springfield}, \textsf{capitalOf}, \textsf{Illinois}). A set of such triples is known as a knowledge base, but can be combined to produce a multi-graph where nodes represent the entities and directed edges represent the predicates. Different predicates can be represented by edge types, resulting in a heterogeneous information network that is often referred to as a \textit{knowledge graph}. Given a knowledge base that is extracted from a large repository of statements, like Wikipedia or the Web at large, the resulting knowledge graph represents \emph{some} of the factual relationships among the entities mentioned in the statements. If there existed an ultimate knowledge graph which knew everything, then fact checking would be as easy as checking for the presence of an edge in the knowledge graph. In reality, knowledge graphs have limited information and are often plagued with missing or incorrect relations making validation difficult.

Although a knowledge graph may be incomplete, we assume that most of the edges in the graph represent true statements of fact. With this assumption, existing fact checking~\cite{Ciampaglia2015} and link prediction methods~\cite{Kleinberg2007,Adamic2003,Barabasi1999,Katz1953,Haveliwala2002} would rate a given statement to be true if it exists as an edge in the knowledge graph or if there is a short path linking its subject to its object, and false otherwise. Statistical relational learning models~\cite{Nickel2011,Socher2013,Bordes2013,Lin2015} can measure the truthfulness by calculating the distance between the entities and predicate in a given statement. However, the limitation of existing models make them unsuitable for fact checking. Link prediction methods, Adamic/Adar~\cite{Adamic2003} and personalized PageRank~\cite{Haveliwala2002} for example, work on untyped graphs and are incapable of capturing the heterogeneity of knowledge graphs; other heterogeneous link prediction algorithms,~\eg, PathSim~\cite{Sun2011} and PCRW~\cite{Lao2010}, not only need human annotated meta paths but also have strict constraints on the input meta paths. Statistical relational learning models such as RESCAL~\cite{Nickel2011}, NTN~\cite{Socher2013}, TransE~\cite{Bordes2013}, and other variants~\cite{Lin2015,Wang2014} utilize the type information in knowledge graphs but can not work with unseen predicate types and do not model the complicated interactions among relations explicitly.

In the present work, we present a discriminative path-based method for fact checking in knowledge graphs that incorporates connectivity, type information, and predicate interactions. Given a statement $\mathcal{S}$ of the form (\textsf{subject}, \textsf{predicate}, \textsf{object}), for example, (\textsf{Chicago}, \textsf{capitalOf}, \textsf{Illinois}), our approach mines discriminative paths that alternatively define the generalized statement (\textsf{U.S. city}, \textsf{predicate}, \textsf{U.S. state}) and uses the mined rules to evaluate the veracity of statement $\mathcal{S}$.

Unlike existing models, the proposed method simulates how experienced human fact-checkers examine a statement: fact-checkers will first attempt to \textit{understand} the generalized notion of the statement using prior knowledge, and then validate the specific statement by applying their knowledge. The statement usually can be generalized by replacing the specific entities by their type-labels. In the present work, we show how to understand a statement by inspecting the related discriminative paths retrieved from the knowledge graph. Returning to the ``\textit{Chicago is the capital of Illinois}'' example, a fact checker, as well as our model, will learn to understand what it means for a \textsf{U.S. city} to be the \textsf{capitalOf} a \textsf{U.S. state}. In this trivial example, a fact checker may come to understand that a city is the capital of a state if the state agencies, governor and legislature are located in the city; from this understanding the fact checker ought to rule that Chicago is not the capital of Illinois because this statement does not satisfy the fact checker's understanding of what \textsf{capitalOf} means.

The advantages of this fact checking procedure is in its {\em generality} and {\em context-dependency}. Just as humans learn unknown words, model generality means the predicate of a statement can be arbitrary and is not required to be presented in the knowledge base. Moreover, once a prior knowledge is learned, it is associated with a certain type of entity pair relation and can be used for different tasks including general question answering or knowledge base completion. 
The notion of context-dependency allows the fact checker to discern different definitions of a predicate in different situations. For example, \textsf{capitalOf} could define the capitals of US states, colloquialisms such as ``Kansas City is the soccer capital of America'', or historical or time-sensitive predicates such as ``Calcutta \emph{was} the capital of India'' depending on the context.


When performed computationally, the task of discovering interesting relationships between or among entities is known generally as association rule mining. Although there has been some effort to adapt association mining for knowledge graph completion, these methods are not well suited for fact-finding and often resort to finding global rules and synonyms~\cite{Abedjan2013,Galarraga2013} rather than generating a robust understanding of the given context dependent predicate~\cite{Meng2015}.

\begin{figure}[t]
\centerline{\includegraphics[width=.73\textwidth]{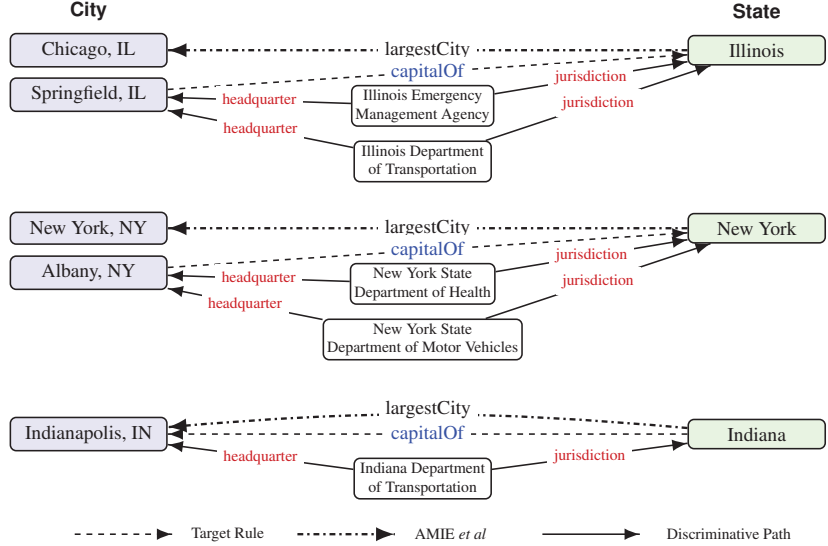}}
\caption{Knowledge graph of US cities and states from DBpedia. \{\textsf{city}\}$\xrightarrow{\textsf{largestCity}^{-1}}$\{\textsf{state}\} and \{\textsf{city}\}$\xrightarrow{\textsf{headquarter}^{-1}}$\{\textsf{entity}\}$\xrightarrow{\textsf{jurisdiction}}$\{\textsf{state}\} are the discriminative paths of \{\textsf{city}\}$\xrightarrow{\textsf{capitalOf}}$\{\textsf{state}\} mined by AMIE~\cite{Galarraga2013} and the proposed method respectively.}
\label{fig:city_capital_example}
\end{figure}

Figure~\ref{fig:city_capital_example} illustrates three graph fragments from the DBpedia knowledge base~\cite{Lehmann2014} containing cities and states. This example demonstrates, via actual results, how the proposed automatic fact checker is able to determine relationships that uniquely define what it means for an entity to be the \textsf{capitalOf} another entity. Association rule miners~\cite{Galarraga2013} and link prediction models~\cite{Barabasi1999,Katz1953} incorrectly indicate that the \textsf{largestCity} is most associated with the \textsf{capitalOf} predicate. In contrast, our framework, indicated by solid edges, finds the rules that most uniquely define what it means to be the \textsf{capitalOf} a state. In this example, our top result indicates that a US state capital is the \textsf{city} in which the \textsf{headquarters} of entities that have \textsf{jurisdiction} in the \textsf{state} are located. In other words, we find that a US state capital is indeed the city where the state agencies, like the Dept. of Transportation, or the Dept. of Health, have their headquarters. 

To summarize, we show that we can leverage a collection of factual statements for automatic fact checking. Based on the principles underlying link prediction, similarity search and network closure, we computationally gauge the truthfulness of an assertion by mining connectivity patterns within a network of factual statements. Our current work focuses on determining the validity of factual assertions from simple, well-formed statements; the related problems of information extraction~\cite{Etzioni2004}, claim identification~\cite{Hassan2015}, answering compound assertions~\cite{Wu2014}, and others~\cite{Nickel2015} are generally built in-support-of or on-top-of this central task.

Recent work in general heterogeneous information networks, of which knowledge graphs are an example, has led to the development of meta path similarity metrics that show excellent results in clustering, classification and recommendation~\cite{Sun2012,Lao2010,Shi2014,Sun2011}. The state of the art in meta path mining works by counting the path-instances or randomly walking over a constrained set of hand-annotated typed-edges~\cite{Sun2011}. Unfortunately, this means that a human has to understand the problem domain and write down relevant meta paths before analysis can begin. In this work, our focus is on methods that automatically determine the set of path-descriptions called \textbf{discriminative paths} that uniquely encapsulate the relationship between two entities in a knowledge graph.



The specific contributions of this paper are as follows:

\begin{enumerate}
\item We developed a fast discriminative path mining algorithm that can discover ``definitions'' of an RDF-style triple, \ie, a statement of fact. The algorithm is able to handle large scale knowledge graphs with millions of nodes and edges.

\item We designed a human interpretable fact checking framework that utilizes discriminative paths to predict the truthfulness of a statement.

\item We modeled fact checking as a link prediction problem and validated our approach on two real world, large scale knowledge graphs, DBpedia~\cite{Lehmann2014} and SemMedDB~\cite{Kilicoglu2012}. The experiments showed that the proposed framework outperforms alternative approaches and has a similar execution time.
\end{enumerate}

In this paper, we incorporate lessons learned from association rule mining and from heterogeneous information network analysis in order to understand the meanings of various relationships, and we use this new framework for fact-checking in knowledge graphs. To describe our approach we first formalize the problem in Sec.~\ref{sec:problem_definition} and define our solution in Sec.~\ref{sec:discriminative_path_discovery}. Section~\ref{sec:experiment} presents extensive experiments on two large, real world knowledge graphs. We present related work in Sec.~\ref{sec:related_work} before drawing conclusions and discussing future work in Sec.~\ref{sec:conclusions}.

\section{Problem Definition} \label{sec:problem_definition}

We view a knowledge graph to be a special case of a heterogeneous information network (HIN) where nodes represent entities and edges represent relationships between entities, and where heterogeneity stems from the fact that nodes and edges have clearly identified type-definitions. The type of an entity is labeled by some ontology, and the type of an edge is labeled by the predicate label. With the above assumptions, we formally define a knowledge graph as follows:

\begin{definition}[Knowledge Graph]A knowledge graph is a directed multigraph $\mathcal{G}=(\mathcal{V},\mathcal{E},\mathcal{R},\mathcal{O}, \psi, \phi)$, where $\mathcal{V}$ is the set of entities, $\mathcal{E}$ is a set of labeled directed edges between 2 entities, $\mathcal{R}$ represents the predicate label set, and $\mathcal{O}$ is the ontology of the entities in $\mathcal{G}$. The ontology mapping function $\psi(v)=\mathbf{o}$, where $v \in \mathcal{V}$ and $\mathbf{o} \subset \mathcal{O}$, links an entity vertex to its label set in the ontology. The predicate mapping function $\phi(e) = p$, where $e \in \mathcal{E}$ and $p \in \mathcal{R}$, maps an edge to its predicate type.
\label{def:knowledge_graph}
\end{definition}

The knowledge graph defined here differs from the standard definition of an HIN; Defn.~\ref{def:knowledge_graph} dictates that a node may be mapped to multiple types, which is unlike the traditional HIN definition in which each node can be mapped to only one type-label~\cite{Sun2011}. When $\psi(v) = \mathbf{o}$ satisfies $|\mathbf{o}| = 1$ for all $v \in \mathcal{V}$, then Defn.~\ref{def:knowledge_graph} degenerates to the standard HIN definition.

For example, the DBpedia knowledge base can be represented as a knowledge graph in which $\mathcal{V}$ represents entities like \textsf{Springfield}, \textsf{Chicago}, or \textsf{Illinois}; $\mathcal{E}$ represents some link between two entities; $\mathcal{O}$ represents a classification scheme like the Wikipedia Category graph with type-labels like \textsf{city} and \textsf{state} categories for \textsf{Chicago} and \textsf{Illinois} respectively; and $\mathcal{R}$ represents the predicate labels like \textsf{capitalOf} and \textsf{largestCity} for edges.

Typed nodes and edges given in the knowledge graph naturally result in an enhanced set of connections called \emph{meta paths} that describe how two entities are connected by their type-labels.

\begin{definition}[Meta Path]Given a knowledge graph $\mathcal{G}=(\mathcal{V},\mathcal{E},\mathcal{R},\mathcal{O}, \psi, \phi)$, a meta path $\Pi^{k}$ is defined as a directed, typed sequence of vertices and edges $\mathbf{o}_1 \xrightarrow{p_1} \mathbf{o}_2 \xrightarrow{p_2} \ldots \xrightarrow{p_{k-1}} \mathbf{o}_k$ in $\mathcal{G}$, where $\mathbf{o}_i$ denotes the set of ontology labels of vertex $i$, $p_i$ represents the predicate of the directed edge that connects vertex $i$ to $i+1$, and $k$ denotes the length of the meta path.
\label{def:meta_path}
\end{definition}

If we relax the definition of a meta path in Defn.~\ref{def:meta_path} in such a way that the edges still carry type-information, but the non-endpoint nodes do not, then the meta path degenerates into an {\em anchored predicate path} anchored by their starting and ending entity-types:

\begin{definition}[Anchored Predicate Path]Given a $k$-length meta path $\Pi^{k}$, the anchored predicate path $P$ is defined as the corresponding directed, typed sequence of edges with typed-endpoints $P^{k} = \textbf{o}_1\xrightarrow{p_1}\xrightarrow{p_2}\ldots\xrightarrow{p_{k-1}}\textbf{o}_k$.
\label{def:anchored_predicate_path}
\end{definition}

With the definitions of knowledge graph and anchored predicate path, here we define the {\em discriminative path} of a statement as:

\begin{definition}[Discriminative Paths]The set of discriminative paths $\mathbf{D}^k_{(\mathbf{o}_u,\mathbf{o}_v)}$ are those anchored predicate paths that alternatively describe the given statement of fact $\mathbf{o}_u\xrightarrow{p}\mathbf{o}_v$, where the maximum path length is $k$.
\label{def:discriminative_path}
\end{definition}  


For example, a meta path $\Pi$ between two entities \textsf{Illinois} and \textsf{Springfield} is represented by the following sequence \{\textsf{city}, \textsf{settlement}\} $\xrightarrow{\textsf{headquarter}^{-1}}$ \{\textsf{state agency}\} $\xrightarrow {\textsf{jurisdiction}}$ \{\textsf{state}\}. The corresponding anchored predicate path $P$ is $\langle$ \textsf{headquarter}$^{-1}$, \textsf{jurisdiction} $\rangle$ anchored by \{\textsf{city}, \textsf{settlement}\} and \{\textsf{state}\}. If $P \in \mathbf{D}$ holds, that means \textsf{capitalOf} can be at least partially defined by $P$. We discuss how to discover these discriminative paths in the next section. 


Note that in our generalization of HIN, the first entity in the meta path is mapped to two type-labels, and could have many more. Entities with many type labels tend to be more prone to label error when using the meta path representation. With this in mind, we choose to use the anchored predicate path representation which we find to have better tolerance on errant type labels. A detailed comparison is given in Sec.~\ref{sec:experiment}.

With the definitions above, the goal of this paper can be formally stated as:


\begin{definition}[Fact Checking]Given a knowledge graph $\mathcal{G}$ and a statement of fact $\mathcal{S}=(s, p, t)$, which may be true or untrue, where subject $s \in \mathcal{V}$, object $t \in \mathcal{V}$. Fact checking is the process of using a learned understanding of the relationship $\mathbf{D}^k_{(\mathbf{o}_u,\mathbf{o}_v)}$ to determine whether the edge $s \xrightarrow{p} t$ is missing in $\mathcal{G}$ such that $\mathbf{o}_s=\mathbf{o}_u$ and $\mathbf{o}_t=\mathbf{o}_v$.
\label{def:fact_checking}
\end{definition}

Simply put, we view the fact checking problem as a supervised link prediction task and validate a proposed fact statement $(s, p, t)$ by determining if that the proposed fact is implied by the data within the knowledge graph. When $p \in \mathcal{R}$ holds, the positive paths $\mathbf{T}^+$ of the predicate $p$ can be automatically generated by $\mathbf{T}^+ = \{(u,v)| u \xrightarrow{p} v \in \mathcal{G}\}$, and negative descriptions  $\mathbf{T}^-$ of the predicate $p$ can be automatically generated by $\mathbf{T}^- = \{(u,v)| u \xrightarrow{p} v \notin \mathcal{G}\}$ such that $\mathbf{o}_u = \mathbf{o}_s$ and $\mathbf{o}_v = \mathbf{o}_t$. 

$\mathbf{T}^+$ and $\mathbf{T}^-$ can also be human provided if $p \notin \mathcal{R}$.

Unlike traditional link prediction problems which simply identify true edges from all possible edges, the fact checking problem is harder because it needs to distinguish true edges $s\xrightarrow{p}t$ from a smaller edge set $\mathbf{\Pi} = \{s^\prime\xrightarrow{p}t^\prime\}$ where $s^\prime$ and $t^\prime$ are have the same type, connectivity, etc. as $s$ and $t$ respectively. Traditional algorithms, which use purely topological features, do not work well in this task because the topology of the network does not sufficiently distinguish between true and false edges.

In this work we propose a model that automatically discovers discriminative paths in order to perform fact checking. The resulting discriminative paths define the proposed predicate $p$ in terms of its subject $s$ and its object $t$ by asking two questions: 1) does the predicate $p$ connect entities that are of the same or similar type as $s$ and $t$ (generality), and 2) do the paths that connect $s$ to $t$ differ from the paths that connect similarly-typed entities but which are \emph{not} connected by $p$ (context-dependent)?


The paths that maximize the above questions 1 and 2 are those paths that uniquely define the predicate in terms of its subject and object. These paths are critical to modelling statements of fact and determining their veracity.

Next we will demonstrate how to test the veracity of statements of fact using discriminative path analysis. 








\section{Discriminative Path Analysis} \label{sec:discriminative_path_discovery}

As defined above, \emph{discriminative} paths $\mathbf{D}^{k}_{(\mathbf{o}_s,\mathbf{o}_t)}$ are those anchored predicate paths that alternatively represent the predicate $p$ from some proposed fact triple $(s, p, t)$ between subjects of the same type as $s$ (denoted $\psi({s})$) and objects of the same type as $t$ (denoted $\psi({t})$). For example, the proposed fact triple at the top of Fig.~\ref{fig:city_capital_example} is (\textsf{Springfield}, \textsf{capitalOf}, \textsf{Illinois}). Other subjects of the same \textsf{\{city, settlement\}}-type include \textsf{Indianapolis}, \textsf{Chicago}, \textsf{New York City}, \textsf{Albany}, etc., and other objects of the same \textsf{\{state\}}-type include \textsf{New York}, \textsf{Indiana}, etc. In this example, one of the predicate paths $P$ that alternatively and uniquely captures this relationship is $\langle$\textsf{headquarter}$^{-1}$,\textsf{jurisdiction}$\rangle$ which is anchored by \{\textsf{city, settlement}\} and \{\textsf{state}\}. 

In addition, $\mathbf{D}^{k}_{(\textbf{o}_s,\textbf{o}_t)}$ includes many other discriminative anchored predicate paths of length $\le k$ that connect $\psi(u)=\textbf{o}_s$ to $\psi(v)=\textbf{o}_t$, \ie, $\{\textsf{city}, \textsf{settlement}\}$ to $\{\textsf{state}\}$.

\begin{figure}
\centerline{\includegraphics[width=\textwidth]{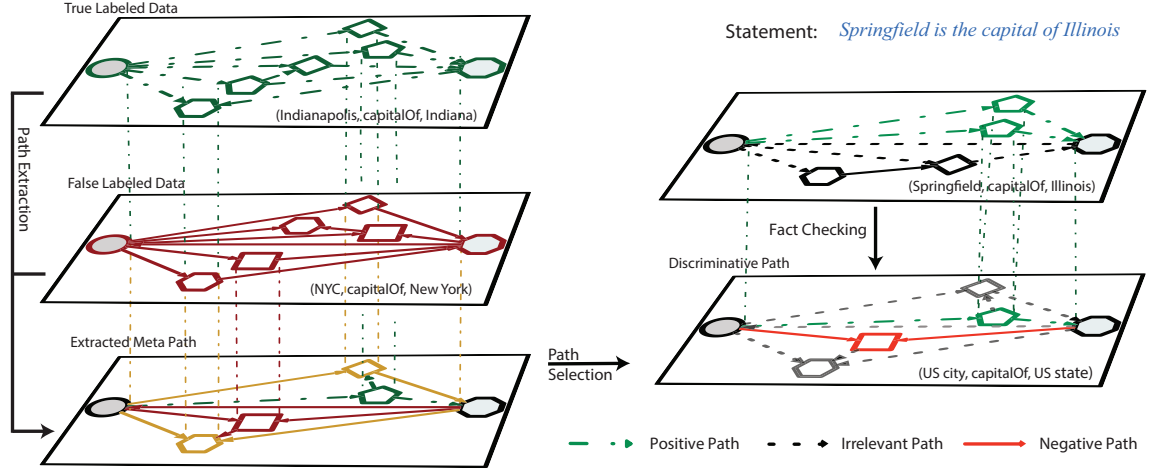}}
    \caption{Overview of the proposed fact checking framework. We first extract meta/predicate paths from labeled data set. Then we preform feature selection on the meta/predicate paths to determine the importance of each meta/predicate path and construct the prediction model. Finally we compare the given statement of fact with the learned model and output the judgement. This figure is best viewed in color.}
    \label{fig:system}
\end{figure}

In Fig.~\ref{fig:system}, we illustrate our proposed fact checking system that contains 3 phases: 1) extraction, 2) selection, and 3) validation. The extraction phase collects anchored predicate paths that alternatively connect the subject and object of a proposed statement of fact. Using the extracted paths $\mathbf{P}^{k}_{(\mathbf{o}_s,\mathbf{o}_t)}$, the selection phase defines the fact with the most discriminating anchored predicate paths $\mathbf{D}^{k}_{(\mathbf{o}_s,\mathbf{o}_t)}$. The validation phase compares the actual statement of fact, \ie, how the subject, predicate and object are actually connected, with a statistical model constructed from the discriminative paths.

\subsection{Path Extraction} \label{sec:meta_path_extraction}

Unlike existing meta path based models, which require hand annotation~\cite{Sun2011,Shi2014} or exhaustive enumeration~\cite{Lao2010} and are impractical on large-scale systems, we learn the best descriptions automatically.

We propose a fast discriminative path discovery algorithm using a constrained graph traversal process. The key idea is the assumption that although the number of paths in a knowledge graph is huge, only a small portion are actually helpful for a given task. Furthermore, among the reduced set of helpful meta paths, only a few may be discriminative in the presence of some predicate.

For example, if (\textsf{Springfield}, \textsf{capitalOf}, \textsf{Illinois}) is the proposed statement of fact in need of checking, the meta paths we are interested in are only those that start at a \textsf{city} and end at a \textsf{state}. So instead of enumerating all possible paths, we collect anchored predicate paths by traversing the graph starting from the given subject-entity and ending at the given object-entity up to a length of $k$.




To find which paths between $\mathbf{o}_u$ and $\mathbf{o}_v$ are most discriminating we further create positive and negative node-pair sets $\mathbf{T}^+$ and $\mathbf{T}^-$ and retrieve two anchored predicate path sets $\mathbf{P}^{+}_{(\mathbf{o}_s,\mathbf{o}_t)}$ and $\mathbf{P}^{-}_{(\mathbf{o}_s,\mathbf{o}_t)}$ respectively using a depth-first multi-graph traversal. Specifically, our DFS-like graph traversal is based on a closure function $\mathbb{C}$, which we define as:

\begin{equation}
\mathbb{C}_p(v) = \left\{(p,v^\prime)|(v,p,v^\prime)\in \mathcal{G}\right\} \cup \left\{(p^{-1},v^\prime)|(v^\prime,p,v)\in \mathcal{G}\right\},
\end{equation}

\noindent{}where $v$ is some entity-node and $p$ denotes the predicate associated with the closure. Simply put, $\mathbb{C}_p(v)$ finds all nodes that can be reached from $v$ via predicate $p$ or $p^{-1}$.

Then we define a transition function $\mathcal{T}(v_i)$ which returns all $v_{i+1}$ candidates, \ie, all of the next entity-nodes, for a path $v_1 \xrightarrow{p_1} v_2 \xrightarrow{p_2} \ldots \xrightarrow{p_{i-1}} v_i$ as 

\begin{equation}
\mathcal{T}(v_i) = \left\{\cup_{p \in \mathcal{R}}\mathbb{C}_{p}(v_{i}) \setminus \cup_{j=1}^{i}\{v_j\}\right\},
\end{equation}

\noindent{}which contains all of the possible next-nodes that can be visited from $\mathbb{C}_p(v)$ except those that have already been visited. Using the closure function $\mathbb{C}_p(v)$ and transition function $\mathcal{T}(v)$, we retrieve the path set $\mathcal{P}$ with path length $\leq k$ by $\mathcal{P} = \cup_{i=1}^k \mathcal{P}^k$, s.t.

\begin{equation}
\begin{split}
\mathcal{P}^k = \{&s, \mathcal{T}(v_1), \mathcal{T}(v_2), \ldots, \mathcal{T}(v_{k-2}), t | \\ & (s,t)\in\mathbf{T}, v_1 = s, v_i \in \mathcal{T}(v_{i-1}), t \in \mathcal{T}(v_{k-1}) \}.
\end{split}
\end{equation}

Unlike traditional graph traversal algorithms that follow the edge direction, our implementation records and follows both directions of each visited edge if possible. In this way, the algorithm actually discovers paths such as \{\textsf{city}\} $\xrightarrow{\textsf{headquarter}^{-1}}$ \{\textsf{state agency}\} $\xrightarrow {\textsf{jurisdiction}}$ \{\textsf{state}\}, which is technically a three node subgraph rather than a path by traditional definitions.

At this point, our framework will have gathered several anchored predicate paths, some of which may be helpful while others may be unhelpful or spurious. Next, we calculate the importance of each extracted anchored predicate path for inclusion into a final regression model.



\subsection{Meta Path versus Predicate Path} \label{sec:metapath_predicatepath}
\ignore{
\begin{table}
    \caption{Example training instance matrix $\mathbf{X}$. Cell values represent the number of the paths in $\mathcal{G}$ anchored by the endpoints of the instances and matching predicate paths in the columns.}

    \resizebox{0.99\textwidth}{!}{
        \begin{tabular}{l || c | c | c | c | l}
    
         & $\langle\textsf{headquarter}^{-1}, \ \textsf{jurisdiction}\rangle$ & $\langle\textsf{location}^{-1}, \ \textsf{jurisdiction}\rangle$ & $\langle\textsf{headquarter}^{-1}, \ \textsf{regionServed}\rangle$ & $\langle\textsf{garrison}^{-1}, \ \textsf{country}\rangle$ & $\mathbf{y}$ \\ \hline
    $(\textsf{Chicago}$, $\textsf{Illinois})$ & 1 & 1 & 1 & 0 & N\\
    $(\textsf{Springfield}$, $\textsf{Illinois})$ & 2 & 2 & 1 & 1 & Y\\
    $(\textsf{Kansas City}$, $\textsf{Kansas})$ & 0 & 0 & 0 & 0 & N \\
    $(\textsf{Albany}$, $\textsf{New York})$ & 2 & 2 & 0 & 0 & Y\\
    $(\textsf{Sacramento}$, $\textsf{California})$ & 10 & 10 & 2 & 2 & Y\\
    $(\textsf{Houston}$, $\textsf{Texas})$ & 0 & 0 & 0 & 0 & N \\
        \end{tabular}
    }
    \label{tab:x_example}
\end{table}
}

Although existing heterogeneous information network analysis methods usually utilize meta paths, in this work we use anchored predicate paths as the feature set. As introduced before, the presence of entity types in meta paths may sometimes be redundant and can typically be inferred by the predicate if there is no ambiguity. For example in the  DBLP network~\cite{Ley2009}, $\textsf{writtenBy}$ always connects a \textsf{paper} with an \textsf{author}, and \textsf{cite} always connects two \textsf{paper}s. In such cases, meta paths can be converted into predicate path without ambiguity.

In more complex knowledge graphs, such as DBpedia~\cite{Lehmann2014} and SemMedDB~\cite{Kilicoglu2012}, an entity can have multiple type-labels. The labels of the same types of entities may not always be consistent due to mislabelling or different interpretations. An example would be the type labels of \textsf{Boston} and \textsf{Sacramento}, which are the capitals of Massachusetts and California respectively. The type-labels of \textsf{Boston} is \{\textsf{city}, \textsf{settlement}, \textsf{populated place}\}, whereas the type-labels of \textsf{Sacramento} are \{\textsf{settlement}, \textsf{populated place}\}. Because the type labels of \textsf{Sacramento} do not exactly match the type-labels of \textsf{Boston}, then a meta path model would treat these paths differently, resulting in many highly overlapping, but not exactly matching, paths. 


We use the Jaccard coefficient~\cite{levandowsky1971distance} to measure the similarity of entity-label sets and use this score to reduce redundant meta paths when possible:
\begin{equation}
 J(\psi(u),\psi(v)) = \frac{|\psi(u) \cap \psi(v)|}{|\psi(u) \cup \psi(v)|}.
 \end{equation}

We combine two meta paths \{\textsf{s}\}$\xrightarrow{p}$\{\textsf{t}\} and \{\textsf{s$^\prime$}\}$\xrightarrow{p}$\{\textsf{t$^\prime$}\} if $J(\psi(s),\psi(s^\prime))$ and $J(\psi(t),\psi(t^\prime))$ are larger or equal to a threshold. In practice, we set the threshold to $0$ for all non-endpoint entities, which means we ignore the entity-type information in meta paths; for meta path endpoints we set the threshold as $\frac{1}{|\psi(u) \cup \psi(v)|}$. Ultimately, this method converts the meta paths into anchored predicate paths.

Intuitively, the use of more tightly constrained meta paths instead of predicate paths ought to increase the information richness leading to better results, but our initial trials showed that: 1) the use of meta paths significantly increased the number of variables in our fact checking model without any noticeable improvement in performance, and 2) the inclusion of noisy entity types from meta paths actually lowered the occurrence-rate of important meta paths resulting in lower discriminative power in the set of paths. A detailed discussion of these counter-intuitive results are provided in the experiment section.

To recap, at this point we have selected positive anchor-entities $\mathbf{T}^+$ and negative anchor-entities $\mathbf{T}^-$. Using those anchors we find many predicate paths $\mathbf{P}^+$ and $\mathbf{P}^-$ that connect the positive and negative anchors; these paths are essentially alternate descriptions of the original predicate that provide evidence that can be used to uniquely define the original predicate.

\ignore{
Because of the issues raised above, in this work we use predicate paths as defined in Defn.~\ref{def:predicate_path} that are anchored by their starting and ending entity-types:

\begin{definition}[Anchored Predicate Path]Given a $k$-length meta path $\Pi^{k}$, the anchored predicate path $P$ is defined as the corresponding directed, typed sequence of edges with typed-endpoints $P^{k} = \textbf{o}_1\xrightarrow{p_1}\xrightarrow{p_2}\ldots\xrightarrow{p_{k-1}}\textbf{o}_k$.
\label{def:anchored_predicate_path}
\end{definition}
}

\subsection{Path Selection} \label{sec:path_selection}

In this section we describe the procedure used to find the most discriminative predicate paths $\mathbf{D}$ from predicate path sets $\mathbf{P}^{+}$ and $\mathbf{P}^{-}$.
For this we define $\mathbf{X}$ to be an $n \times m$ training instance matrix, wherein the $i^{\textrm{th}}$ row in $\mathbf{X}$ represents a training instance from a pair of anchors $u$ and $v$ such that $\mathbf{o}_u=\mathbf{o}_s$ and $\mathbf{o}_v=\mathbf{o}_t$. The cell $\mathbf{X}_{i,j}$ is the number of anchored predicate paths $P_j$ that are anchored by $u$ and $v$. Class labels indicate if the training instance is connected by the predicate of interest or not. The goal of path selection is to create a new $n \times m^\prime$ matrix $\mathbf{X^\prime}$, where $m^\prime$ contains only the paths/features with the most discriminative power. This is achieved by a feature selection function:


\begin{equation}
\begin{split}
\mathbf{X}^{\prime} = f(\mathbf{X},\mathbf{w},\delta) = \mathbf{X}_{1:n, \{ j | j \in 1:m, w_j \geq \delta \}},
\end{split}
\end{equation}

\noindent where $\mathbf{w}$ is an $m$-dimensional feature importance vector, and $\delta$ is an importance threshold. 

The importance $w_j\in\mathbf{w}$ of a predicate path $P_j\in\mathbf{P}$ is measured using the information gain of $\mathbf{X}_{:,j}$ and $\mathbf{y}$:

\begin{equation}
I(\mathbf{X}_{:,j}:\mathbf{y}) =
\sum_{x_{i,j} \in \mathbf{X}_{:,j}} \sum_{y_i \in \mathbf{y}} p(x_{i,j})p(y_i) \log \left (\frac{p(x_{i,j},y_i)}{p(x_{i,j})p(y_{i})} \right ),
\end{equation}

\noindent where $\mathbf{X}_{:,j}$ denotes the column vector of feature $j$, $\mathbf{y}$ represents the corresponding label vector, and $x_{i,j}$ is the data value of the cell at $\mathbf{X}_{i,j}$~\cite{Quinlan2014}. In order to reduce the rank of $\mathbf{X}$ we set $\delta$ empirically. 


With the discriminative predicate paths extracted, pruned and represented in $\mathbf{X}^\prime$, we train a standard logistic regression model and use it to validate the original statement of fact. 

\ignore{
\begin{equation}
\begin{split}
    p(y_i = 1) = \frac{e^{\beta_0 + \mathbf{X}^\prime_{i,:}\beta}}{1 + e^{\beta_0 + \mathbf{X}^\prime_{i,:}\beta}} \ ,
\end{split}
\end{equation}

\noindent where $\mathbf{X}^\prime_{i,:}$ and $y_i$ denotes the input feature vector and the class label of instance $i$ respectively, and $\beta$ represents the learned regression coefficients~\cite{Murphy2012}.
}

\begin{table}[t]
    \caption{Top discriminative paths defining \textsf{capitalOf}, ordered by $\mathbf{w}$.}

    \begin{tabular}{c | c  }
    Rank  & \textbf{Meta Path $\mathbf{\Pi}$}  \\
    \hline\hline
    1 & $\{\textsf{city},\textsf{settlement}\} \xrightarrow{\textsf{location}^{-1}} \{\textsf{state agency}\} \xrightarrow {\textsf{location}} \{\textsf{state}\}$ \\
    
    2 & $\{\textsf{city},\textsf{settlement}\} \xrightarrow{\textsf{deathPlace}^{-1}} \{\textsf{person}\} \xrightarrow {\textsf{deathPlace}} \{\textsf{state}\}$ \\
    
    3 & $\{\textsf{city},\textsf{settlement}\} \xrightarrow{\textsf{headquarter}^{-1}} \{\textsf{state agency}\} \xrightarrow {\textsf{jurisdiction}} \{\textsf{state}\}$ \\
    
    4 & $\{\textsf{city},\textsf{settlement}\} \xrightarrow{\textsf{location}^{-1}} \{\textsf{state agency}\} \xrightarrow {\textsf{jurisdiction}} \{\textsf{state}\}$  \\
    
    5 & $\{\textsf{settlement}\} \xrightarrow{\textsf{location}^{-1}} \{\textsf{state agency}\} \xrightarrow {\textsf{jurisdiction}} \{\textsf{state}\}$ \\\hline

\multicolumn{1}{c}{} & \textbf{Anchored Predicate Path $\mathbf{D}$} \\ \hline\hline

1 & $\langle\textsf{headquarter}^{-1}, \ \textsf{jurisdiction}\rangle$ \\

2 & $\langle\textsf{location}^{-1}, \ \textsf{jurisdiction}\rangle$ \\

3 & $\langle\textsf{headquarter}^{-1}, \ \textsf{regionServed}\rangle$ \\
 
4 & $\langle\textsf{garrison}^{-1}, \ \textsf{country}\rangle$ \\

5 & $\langle\textsf{deathPlace}^{-1}, \ \textsf{deathPlace}\rangle$ \\\hline

\multicolumn{1}{c}{} & \textbf{Discriminative Anchored Predicate Path $\mathbf{D}^*$} \\ \hline\hline

1 & $\langle\textsf{headquarter}^{-1}, \ \textsf{jurisdiction}\rangle$\\
2 & $\langle\textsf{location}^{-1}, \ \textsf{jurisdiction}\rangle$\\
3 & $\langle\textsf{garrison}^{-1}, \ \textsf{country}\rangle$\\
4 & $\langle\textsf{headquarter}^{-1}, \ \textsf{parentOrganisation}\rangle$ \\
5 & $\langle\textsf{location}^{-1}, \ \textsf{parentOrganisation}\rangle$ \\\hline
    \end{tabular}
\label{tab:disc_path}
\end{table}


\subsection{Fact Interpretation} \label{sec:fact_interpertation}


Although the discriminative paths are used for fact checking, there is no guarantee that all of the predicate paths actually describe intuitive or important attributes of the given statement of fact. For example, the fact checking model trained with the statement of fact (\textsf{Springfield}, \textsf{capitalOf}, \textsf{Illinois}) contains many spurious predicate paths like $\langle\textsf{location}^{-1}, \textsf{location}\rangle$ and $\langle\textsf{deathPlace}^{-1}, \textsf{deathPlace}\rangle$. These predicate paths indicate that ``the capital is located in the state'' and that ``a city is the death place of a person who died in that state'', which are indeed accurate in their respective instances, but not very descriptive of what \textsf{capitalOf} actually means. Although these supportive predicate paths seem superfluous they may actually be used to help identify false statements, according on their learned regression weights. However, these spurious statements should probably not be included in any ``definition'' of a given fact; instead, we want {\em human-interpretable} definitions to consist of only the most important predicates. In other words, we need to find those important discriminative predicate paths that define \emph{only} the predicate in question.



To do this, we sort all of the extracted predicates by their importance $\mathbf{w}$ and construct an ordered list $P_x \prec P_y \prec \ldots$, where $w_x \geq w_y$.

After ordering the predicate paths, we remove unnecessary and verbose predicate paths by
\begin{equation}
\mathbf{D}^* = \{ P | P \in \mathbf{D} \setminus \{ P_j | P_j \in \mathbf{P}^-, \sum_{i=0}^{i=n}\mathbf{X}_{i,j} \geq \theta \} \},
\end{equation}

\noindent where $\theta$ is an importance threshold chosen empirically and varies between 10 and 20. As a result of this function, the set of important discriminative predicate paths $\mathbf{D}^*$ contains the specific definers of the provided predicate. The top 5 discriminative predicate paths for \textsf{capitalOf} are illustrated in Table~\ref{tab:disc_path}.


\section{Experiments} \label{sec:experiment}

In this section we report the results of two tasks: 1) fact checking and 2) definition interpretation using thousands of fact statements from eight different test cases on two large, real world knowledge graphs: DBpedia~\cite{Lehmann2014} and SemMedDB~\cite{Kilicoglu2012}. Before we present the results, we describe the datasets, alternative approaches and the experimental setup.

\begin{table}[t]
    \label{tab:stat}
    \centering
    \caption{Statistics of knowledge graph datasets.}
        \begin{tabular}{c|r r r r r r}
         \multicolumn{1}{c|}{$\mathcal{G}$} & \multicolumn{1}{c}{$|\mathcal{V}|$} & \multicolumn{1}{c}{$|\mathcal{E}|$} & \multicolumn{1}{c}{$|\mathcal{R}|$} & \multicolumn{1}{c}{$|\mathcal{O}|$} & \multicolumn{1}{c}{$|\mathbf{o}| = 0$} & \multicolumn{1}{c}{$|\mathbf{o}| > 1$}\\\hline
         DBpedia~\cite{Lehmann2014}   & 4,743,012 & 27,313,477 & 671 & 451 & 524,889 & 3,313,257 \\
         SemMedDB~\cite{Kilicoglu2012}  & 282,934   & 82,239,653 & 58  & 132 &     137 & 73,936 \\ 
        \end{tabular}
\end{table}

\subsection{Data Set}

The fact checking model requires a knowledge graph as input. For these experiments we generated two large heterogeneous information networks from two widely used knowledge bases. In order to construct a heterogeneous multigraph from each knowledge base, we converted each RDF triple (\textsf{subject}, \textsf{predicate}, \textsf{object}) into a directed edge $\textsf{subject} \xrightarrow{\textsf{predicate}} \textsf{object}$ in $\mathcal{G}$, we further combined entities with same name or identifier into a single entity node in the final knowledge graph. The statistics of two resulting knowledge graphs are shown in Table~\ref{tab:stat}.

\vspace{5pt}\noindent{\textbf{DBpedia}}. DBpedia is a community project which converts facts extracted from Wikipedia into knowledge base triples that follow Semantic Web and Linked Data standards. The resultant knowledge base is split into several components such as infobox-facts (\ie, $(s,p,t)$), and entity type mappings (\ie, $\psi(u) = \mathbf{o}_u$).

From this knowledge base, we use the infobox-facts and article ontology from the April 2014 DBpedia knowledge base to create nodes and edges. We did not include article content, such as text and hyperlinks, in the knowledge graph because it was not the focus of this work.


\vspace{5pt}\noindent{\textbf{SemMedDB}}. The Semantic MEDLINE Database contains $82$ million triples extracted from biomedical text using the SemRep extractor~\cite{Rindflesch2003}. Unlike DBpedia, which does not have duplicate triples, SemMedDB contains a large number of duplicate records and uses the amount of duplication as a measure of credibility. For example, an incorrect statement (\textsf{Chicago}, \textsf{isA}, \textsf{country}) appears only once in the SemMedDB knowledge base, while the correct statement (\textsf{Chicago}, \textsf{isA}, \textsf{city}) appears 10 times. Interestingly, although there are $82$ million edges in SemMedDB, only $20.9\%$ of the edges are unique.

We use the June 2012 version of SemMedDB and translate it to a knowledge graph in the same way as with DBpedia. We do not remove any duplicate edges because comparable algorithms often work better on multigraphs; Adamic/Adar, for example, may leverage duplicate edge information to improve accuracy. 


\subsection{Experiment Setting}

We view the fact checking task as a type of link prediction problem because a fact statement $(s,p,t)$ can be naturally considered as an edge $s\xrightarrow{p}t$ in a given knowledge graph $\mathcal{G}$. The probability that an unknown statement of fact $s\xrightarrow{p}t$ is true is equivalent to the probability that the edge $s\xrightarrow{p}t$ is missing in $\mathcal{G}$. To test the ability of our method to validate missing facts and unseen relations, we remove all edges labelled by the given predicate $p$ and perform fact checking on the modified multigraph $\mathcal{G}^\prime=\mathcal{G}-p$.

All experiments are performed using 10-fold cross validation. The source code of our method and the comparison algorithms, including data preprocessing tools, can be found at \url{https://github.com/nddsg/KGMiner}.

We compared our fact checking algorithm with 9 alternative approaches including Adamic/Adar (AA)~\cite{Adamic2003}, Preferential Attachment (PA)~\cite{Barabasi1999}, Katz~\cite{Katz1953} with $k=3$ and $\beta=0.05$ as recommended by Kleinberg and Liben-Nowell~\cite{Kleinberg2007}, Semantic Proximity (SP)~\cite{Ciampaglia2015} with $k = 3$, personalized PageRank (PPR)~\cite{Haveliwala2002} with damping factor $d=0.15$, SimRank~\cite{Jeh2002}, Path-Constrained Random Walk (PCRW)~\cite{Lao2010}, AMIE~\cite{Galarraga2013}, and TransE~\cite{Bordes2013}.

\ignore{
For a given node pair $(u,v)$, we calculate the scores of aforementioned methods as follows:
$$score_{aa}(u,v) =\sum_{\{x | (u,x)\in \mathcal{E}, (x,v) \in \mathcal{E}\}}\frac{1}{log(|\Gamma(x)|)},$$
$$score_{pa}(u,v) =|\Gamma(u)| \times |\Gamma(v)|,$$
$$score_{katz}(u,v) =\sum_{i=1}^{k}\beta^i|\textrm{path}^i{(u,v)}|,$$
$$score_{sp}(u,v) = \max \left( \frac{1}{1 + \sum_{i=2}^{k-1}log(|\Gamma(x_i)|)} \right),$$
$$score_{ppr}(u,v) = d\delta_{u,v} + (1-d)\sum_{(x,v)\in\mathcal{E}}\frac{score_{ppr}(u,x)}{|\Gamma^+(x)|},$$

\noindent{}where $\Gamma(x)$ denotes the neighborhood set of an entity $x$, $k$ represents the path length, and is set to $k=3$ throughout these experiments; $\beta$ is a parameter $> 0$ that we set to $\beta=0.05$ as recommended by Kleinberg and Liben-Nowell~\cite{Kleinberg2007}; $\textrm{path}^i(u,v)$ is a set of homogeneous paths connecting $u\leadsto v$ within length $i$; $d$ represents the PageRank damping factor, and is set to $d=0.15$, $\delta_{u,v}$ is the indicator function of $u=v$, and $\Gamma^+(x)$ denotes out-going neighbors of $x$.
}
In order to run SimRank on the large knowledge graphs, we implemented Kusumoto and Maehara's SimRank approximation~\cite{Kusumoto2014} with $c=0.8$, $T=100$ and $R=10^4$ set according to the values in their original work.

We use Lin~\etal's TransE implementation~\cite{Lin2015} with $\lambda=0.01$, $\gamma=1$, and $d=L_2$ according to Bordes~\etal~\cite{Bordes2013}. The feature dimension is set to $100$ and the training phrase stops after $1,000$ epochs.

Although AMIE is designed for association rule mining, in this work we employ it as a link prediction method by assuming that the given statement of fact $(s,p,t)$ is true if and only if at least one association rule found by AMIE connects $s$ and $t$ in the graph $\mathcal{G}^\prime$. 

As for meta path based methods, such as PCRW, we use the association rule mined by AMIE as the input rather than hand-labeled meta paths. We chose to use AMIE instead of the discovered meta paths from Meng {\em et al.}~\cite{Meng2015} because the implementation of AMIE is publicly available. Unfortunately, PathSim~\cite{Sun2011} requires input meta paths to be symmetric, \ie, $a\rightarrow b\rightarrow a$ or $a\rightarrow b\rightarrow c\rightarrow b\rightarrow a$, but the rules mined by AMIE are very rarely symmetric in our test cases because the $(s,p,t)$ endpoints typically have different entity-type labels; therefore we cannot compare our method with PathSim and other symmetric-only algorithms.

Due to the large size of the knowledge graphs, it is impractical to run AMIE to completion. In these experiments, we executed AMIE for $2,690$ CPU hours on DBpedia and $1,190$ CPU hours on SemMedDB. The number of AMIE-mined rules on the knowledge graphs is $1,326$ and $5,188$ respectively.

We also tried other statistical relational learning models including RESCAL~\cite{Nickel2011} and NTN~\cite{Socher2013} but the publicly available implementations were either incapable of dealing with the huge data sets we use in this work or returned incomprehensible results. We do not compare search engine based models~\cite{Li2011} because, unlike the original authors, we do not have access to search engine APIs to the extent necessary to carry out a proper comparison.

\subsection{Test Cases} \label{sec:test_case}

Here we briefly describe the test cases we use for fact checking. Each test case is constructed to be as difficult as possible. 

\vspace{5pt}\noindent{\textbf{CapitalOf \#1}}. Check the capital of a US state. In this task we check $\{\textsf{city}\}\xrightarrow{\textsf{capitalOf}}\{\textsf{state}\}$ for the top 5 most populous cities in all 50 US states. In 9 instances the capital city is not in the set of top 5 most populous cities of a state, in these cases we further include the capital city in the test set thereby checking a total of $5\times 50 + 9=259$ statements of fact with 50 true instances, and 209 false instances.

\vspace{5pt}\noindent{\textbf{CapitalOf \#2}}. Check the capital of a US state. In this task we check $\{\textsf{city}\}\xrightarrow{\textsf{capitalOf}}\{\textsf{state}\}$ by creating 200 incorrect random matchings of capitals to states. For example, we check if Springfield, the actual capital of Illinois, is the capital of 4 other states. This random assignment results 250 statements of fact with 50 true instances and 200 false instances.

\vspace{5pt}\noindent{\textbf{US Civil War}}. Check the commander of a US Civil War battle. In this task we check $\{\textsf{person}\}\xrightarrow{\textsf{commanderOf}}\{\textsf{battle}\}$ by creating 584 incorrect random matchings of civil war commanders to civil war battles, as well as 126 true statements about the Union and Confederate commanders of Class A (\ie, decisive) US Civil War battles. 

\vspace{5pt}\noindent{\textbf{Company CEO}}. Check CEO of a company. In this task we check $\{\textsf{person}\}\xrightarrow{\textsf{keyPerson}}\{\textsf{company}\}$ by creating 1025 incorrect random matchings of CEOs to companies, as well as true statements about the CEOs of the 205 notable companies in the Wikipedia \textsf{List of chief executive officers}. 

\vspace{5pt}\noindent{\textbf{NYT Bestseller}}. Check author of a book. In this task we check $\{\textsf{person}\}\xrightarrow{\textsf{author}}\{\textsf{book}\}$ by creating 465 incorrect random pairs of authors to books, as well as true statements about the 63 authors who wrote 93 books that appeared on the New York Times bestseller list between 2010-2015.

\vspace{5pt}\noindent{\textbf{US Vice-President}}. Check vice president of a president. In this task we check $\{\textsf{person}\}$ $\xrightarrow{\textsf{vicePresidentOf}} \{\textsf{person}\}$ by creating 227 incorrect pairs of vice-presidents to presidents, as well as true statements about the 47 vice-presidents of the United States.

\vspace{5pt}\noindent{\textbf{Disease}}. Check if amino acid, peptide, or protein causes a disease or syndrome. In this task we check $\{\textsf{aapp}\}\xrightarrow{\textsf{causes}}\{\textsf{dsyn}\}$ where \textsf{aapp} and \textsf{dsyn} are types in SemMedDB corresponding to amino acid, peptide, or protein and disease or syndrome respectively. We do this by creating 457 incorrect statements, as well as 100 true statements.

\vspace{5pt}\noindent{\textbf{Cell}}. Check if a gene causes a certain cell function. In this task we check $\{\textsf{gngm}\}\xrightarrow{\textsf{causes}}\{\textsf{celf}\}$ where \textsf{gngm} and \textsf{celf} are types in SemMedDB corresponding to gene or genome and cell function respectively. We do this by creating 435 incorrect statements, as well as 99 true statements.

These eight test cases listed above represent a $20/80$ true to false label split of instances. We will experiment with different label ratios in later experiments.

\subsection{Predicate Path Analysis}

\ignore{
\begin{table}[t]
    \centering
    \caption{AUROC score of fact checking on FB15k. The higher the better.}
    \label{tab:fb15}
\resizebox{\textwidth}{!}{
    \begin{tabular}{l | c c c c c c c c c c}
    \hline
     & nomination/nominee & film/release region & nomination/award & award/nominee & person/profession & profession/person & actor/film & film/actor & nominated for & nomination/award \\
    \hline
    TransE   & $0.989$ & $0.691$ & $0.951$ & $0.937$ & $0.940$ & $0.777$ & $0.940$ & $0.950$ & $0.855$ & $0.911$ \\
    PredPath & $0.977$ & $0.675$ & $0.929$ & $0.926$ & $0.872$ & $0.780$ & $0.951$ & $0.953$ & $0.859$ & $0.875$ \\
    \end{tabular}
}
\end{table}
}

\begin{figure}[t]

\centerline{\includegraphics[width=.95\textwidth]{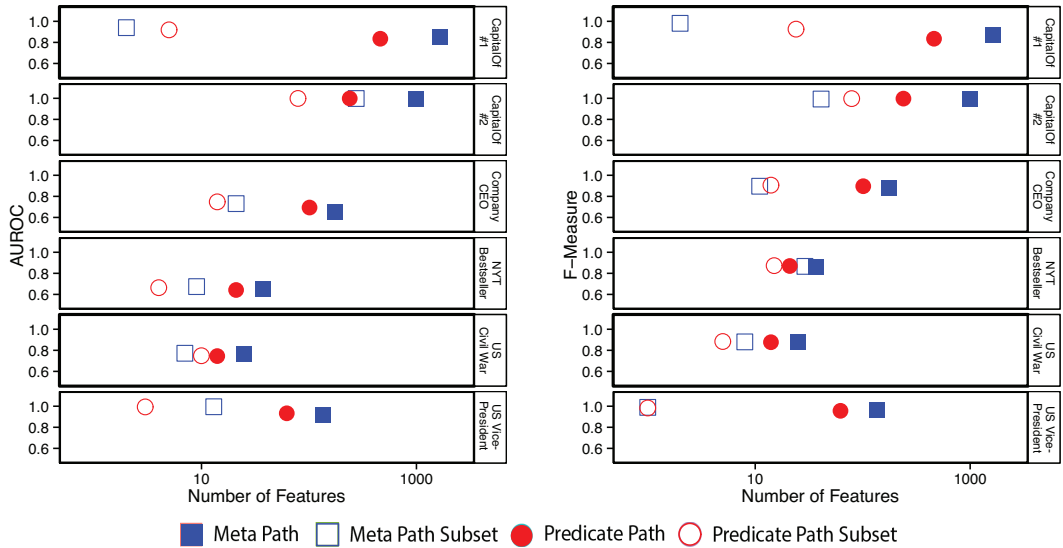}}

    \caption{Fact-checking performance on DBpedia. Solid and hollow symbols denote original and selected best performance subset respectively. This figure demonstrates that anchored predicate paths contain fewer features but have similar performance compared to sets of meta paths. Top left is better.}
    \label{fig:metapath_predicatepath}
\end{figure}

Earlier we argued in favor of anchored predicate paths over the use of meta paths. Recall that the reasoning behind this choice is that the anchored predicate paths are less restrictive than meta paths, which require one or more type-labels for every entity in the path, whereas anchored predicate paths only require type-labels for entities on the endpoints of the path.



Figure~\ref{fig:metapath_predicatepath} shows the results of a comparison between meta paths $\mathbf{\Pi}$ and discriminative anchored predicate paths $\mathbf{D}$ on the 6 DBpedia tasks. We find that the performance of anchored predicate paths (solid circle) are comparable to meta paths (solid square) despite having a much smaller feature set.

We also constructed a subset of meta paths and anchored predicate paths by computing the information gain of each path, sorting by the information gain and chosing the top $k$ that maximizes the area under the receiver operator characteristic (AUROC) score. The empirical result of the meta path subset (hollow square) and anchored predicate path subset (hollow circle) is also illustrated in Fig.~\ref{fig:metapath_predicatepath}. 

We find that, even if we select the most informative paths, the feature set generated by meta paths is typically bigger than the set of predicate paths, but results in similar performance. Moreover, the $165,331$ unique meta paths extracted from SemMedDB that match (\textsf{gngm},\textsf{causes},\textsf{celf}) was too large to work with effectively. On the other hand, the number of unique anchored predicate paths totalled only $1,066$.

Apart from the increase in feature set size, the use of meta paths also reduced the understandability of the results. The top discriminative paths found from the \{\textsf{city}\} $\xrightarrow{\textsf{capitalOf}}$ \{\textsf{state}\} example in Table~\ref{tab:disc_path} from earlier show that the re-ranked anchored predicate paths $\mathbf{D}^*$ are more intuitive than the discovered meta paths $\mathbf{\Pi}$. Unfortunately, ``intutive''-ness is a difficult concept to test fully, so we leave a complete test of the understandability of predicate paths and meta paths as a matter for future work.

    \clearpage
    
\begin{table}[t]
    \centering
    \caption{Result of fact checking test cases. The score is the area under ROC curve score computed by logistic regression with 10-fold cross validation. $^*$ means the value is missing due to the large size of feature set (Section~\ref{sec:metapath_predicatepath}). All test cases are explained in Sec.~\ref{sec:test_case}.}
    \label{tab:datasets}
\resizebox{\linewidth}{!}{
    \begin{tabular}{l | c c c c c c c c}
    \hline
        Algorithm  & CapitalOf \#1 & CapitalOf \#2 & Company CEO & NYT Bestseller & US Civil War & US Vice-President & Disease & Cell\\ 
    \hline
        Adamic/Adar~\cite{Adamic2003}  & $0.387$ & $0.962$  & $0.665$ & $0.650$ & $0.642$ & $0.795$ & $0.671$ & $0.755$\\
        Semantic Proximity~\cite{Ciampaglia2015} & $0.706$ & $0.978$ & $0.614$ & $0.641$ & $0.582$ & $0.805$ & $0.871$ & $0.840$\\
        Preferential Attachment~\cite{Barabasi1999} & $0.396$ & $0.516$ & $0.498$ & $0.526$ & $0.599$ & $0.474$ & $0.563$ & $0.755$\\
        Katz~\cite{Katz1953} & $0.370$ & $0.976$ & $0.600$ & $0.623$ & $0.585$ & $0.791$ & $0.763$ & $0.832$ \\
        SimRank~\cite{Jeh2002} & $0.553$ & $0.976$ & $\mathbf{0.824}$ & $\mathbf{0.695}$ & $0.685$ & $0.912$ & $0.809$ & $0.749$ \\
        AMIE~\cite{Galarraga2013} & $0.550$ & $0.500$ & $0.669$ & $0.520$ & $0.659$ & $0.987$ & $0.889$ & $0.898$ \\
        Personalized PageRank~\cite{Haveliwala2002}  & $0.535$ & $0.535$ & $0.579$ & $0.529$ & $0.488$ & $0.683$ & $0.827$ & $0.885$ \\
        Path-Constrained Random Walk~\cite{Lao2010} & $0.550$ & $0.500$ & $0.542$ & $0.486$ & $0.488$ & $0.672$ & $0.911$ & $0.765$ \\
        TransE~\cite{Bordes2013} & $0.655$ & $0.775$ & $0.728$ & $0.601$ & $0.612$ & $0.520$ & $0.532$ & $0.620$ \\
    \hline
        \textbf{Discriminative Predicate Path ($\mathbf{D}$) Count} & $0.920$ & $\mathbf{0.999}$ & ${0.747}$ & $0.664$ & $0.749$ & $0.993$ & $\mathbf{0.941}$ & $\mathbf{0.928}$\\
        \textbf{Discriminative Meta Path ($\mathbf{\Pi}$) Count} & $\mathbf{0.940}$ & $0.998$ & $0.731$ & $0.674$ & $\mathbf{0.772}$ & $\mathbf{0.995}$ & $*$ & $*$ \\
    \hline
    \end{tabular}
}
\end{table}

\begin{table}[t]
\centering
\caption{Seconds each algorithm consumes for feature generation. The value represents the average feature generation time of each statement. The execution time of AMIE and TransE includes the average time spent on association rule mining and embedding learning respectively. PredPath is faster than other typed methods.}
\label{tab:time}
\resizebox{\linewidth}{!}{
\begin{tabular}{l | c c c c c c c c c c c}
    \hline
    Model & Adamic/Adar & AMIE & Katz & PA & PCRW & PPR & PredPath & Semantic Proximity & SimRank & TransE \\
    \hline
    Time(second) & $0.04$     & $3194.88$ & $0.57$ & $0.03$ & $77.16$ & $446.95$ & $1.00$ & $0.94$ & $0.86$ & $368.32$\\
\end{tabular}
}
\end{table}


\subsection{Fact Checking}

\begin{figure}[t]
    \centerline{\includegraphics[width=\textwidth]{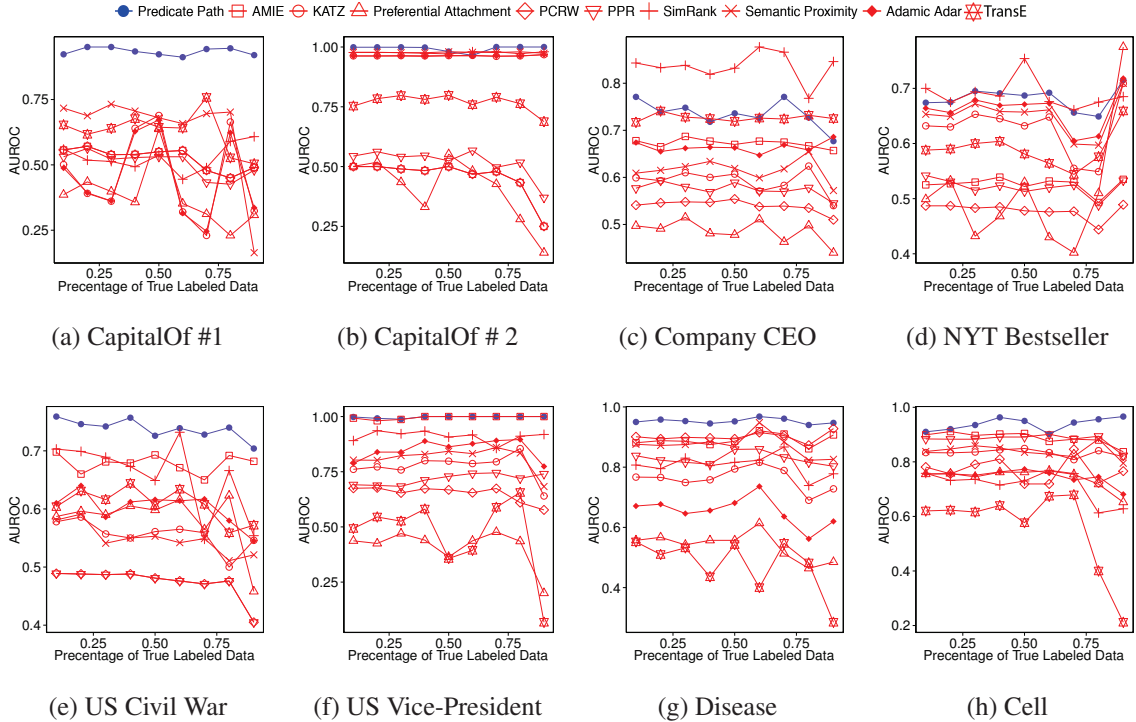}}
    \caption{Fact checking performance with different true to false ratio of labeled data.}
    \label{fig:robustness}
\end{figure}

In Table~\ref{tab:datasets} we compare the proposed fact-checking algorithm with 9 other link prediction, knowledge base completion and data mining algorithms on data from DBpedia and SemMedDB.

In the \textbf{CapitalOf \#1} task, where the true \textsf{capitalOf} statements are mixed with false statements that match the \textsf{largestCity} predicate, the proposed method is shown to significantly outperform other methods. Recall that the set of discriminative predicate paths represent a sort of ``definition'' of the given predicate that is used as a model for fact checking. The top 5 most discriminative predicate paths for the \textbf{CapitalOf \#1} task were originally shown in Table~\ref{tab:disc_path}, and the top 2 most discriminative predicate paths are shown in Table~\ref{tab:missed_path}.

The Adamic/Adar, Preferential Attachment, and Katz models performed very poorly in this example because the features that these purely topographical models rely on most strongly connects the largest city with the state. Unfortunately, only 17 US capital-cities are also the largest city in their state resulting in very poor performance for the topographical models. 


Tasks in which the negatively-labeled data is randomly generated, as in \textbf{CapitalOf \#2} for example, are easier for topological models because, in many cases, the true-labeled statement is the one that is the best connected (especially when compared to random statements). Interestingly, SimRank performs slightly better than our model on the \textbf{Company CEO} and \textbf{NYT Bestseller} tasks. This is most probably because of the high connectivity between the path anchors, and because of the lack of meta path variation, \eg, book authors and company CEOs have relatively few alternate paths that are suitable defining the given statement.



Despite being a deep learning, word2vec-like knowledge base completion system, TransE does not perform well in these tasks. This may due to the large knowledge graph we used in this work, but may also be because TransE is not designed to accept duplicated edges, which seems to help identify factual relations especially in the SemMedDB dataset.

Recall that these results use a true/false label ratio of $20/80$ to simulate real-world fact checking scenarios where the proportion of false statements are significantly larger than true statements. This is not to say that there are more false statements in real-life, just that there are many more possible false statements than there are true statements. With this in mind, we further test the robustness of our model under various true/false label proportions. Figure~\ref{fig:robustness} illustrates the results of these tests where the discriminative predicate path performance (solid blue circle) is found to be relatively invariant to the percentage of labeled data as it changes from 10\% positively-labeled to 90\% positively labeled.

Apart from the accuracy and robustness tests above, we also analyze the amount of time that each algorithm uses while calculating the score for a single statement of fact, \ie, the time it takes to calculate $\mathbf{X}^\prime$. The 8 tasks have similar computational complexity, so we combine the execution times and present the mean average in Table~\ref{tab:time}. We find that our method (labeled PredPath), although slower than shared neighbor methods and untyped path based model such as Adamic/Adar, Preferential Attachment, SimRank and Katz, has an execution time comparable to heterogeneous path-based method Semantic Proximity, and is faster than stochastic models like TransE, Path Constrained Random Walk (PCRW), personalized PageRank (PPR), and the fast, approximate version of SimRank.  


\ignore{
\begin{figure}[t]
\centering
\includegraphics[width=0.65\textwidth]{./figs/time_wide}
\caption{Time consumption of each algorithm. Point represents the average feature generation time of one query. Error bars represent 95\% confidence interval over the 8 tasks. Lower is better. The execution time of AMIE does not include the time ($\approx 4,000$ hours in total) spent on association rule mining.}
\label{fig:time}
\end{figure}
}

\subsection{Statement Interpretation}

\begin{table}[t]
    \centering
    \caption{Top discriminative paths found by proposed method that are missing in AMIE. Predicate path anchors are for illustrative purposes and do not represent the full entity label set.}
    \resizebox{0.75\textwidth}{!}{
        \begin{tabular}{c|r c l}
        \hline
        Task & & Top Discriminative Path Missed by AMIE &\\
        \hline
        CapitalOf \#1 & \{\textsf{city}\} & \begin{tabular}{c}
                $\langle\textsf{headquarter}^{-1}, \textsf{jurisdiction}\rangle$\\
                $\langle\textsf{location}^{-1}, \textsf{jurisdiction}\rangle$ \\
                \end{tabular} & \{\textsf{state}\} \\
        \hline
        CapitalOf \#2 & \{\textsf{city}\} & \begin{tabular}{c}
                    $\langle\textsf{location}^{-1},\textsf{location}\rangle$ \\
                    $\langle\textsf{isPartOf}\rangle$ \\
                    \end{tabular} & \{\textsf{state}\} \\
        \hline
        Company CEO & \{\textsf{person}\} & \begin{tabular}{c}
                $\langle\textsf{employer}\rangle$\\
            \end{tabular} & \{\textsf{company}\} \\
        \hline
        US Civil War & \{\textsf{person}\} & \begin{tabular}{c}
                        $\langle\textsf{notable commander}^{-1},\textsf{takePartIn}\rangle$ \\
                    \end{tabular} & \{\textsf{battle}\}\\
        \hline
        NYT Bestseller & \{\textsf{person}\} & \begin{tabular}{c c}
                            $\langle\textsf{notable work},\textsf{previous work}\rangle$ \\ 
                            $\langle\textsf{notable work},\textsf{subsequent work}\rangle$ \\
                         \end{tabular} & \{\textsf{book}\} \\ 
        \hline
        US President   & \{\textsf{vice president}\} & \begin{tabular}{c}
                                $\langle\textsf{successor},\textsf{president}^{-1}\rangle$ \\
                          \end{tabular} & \{\textsf{president}\}\\ 
       \hline
       Disease & \{\textsf{aapp}\} & \begin{tabular}{c}
                                $\langle\textsf{associatedWith},\textsf{isA}\rangle$ \\
                                $\langle\textsf{stimulates},\textsf{affects}\rangle$ \\
                          \end{tabular} & \{\textsf{dsyn}\} \\ 
                          
     \hline
      Cell & \{\textsf{gngm}\} & \begin{tabular}{c}
                $\langle\textsf{comparedWith},\textsf{negativeAssociatedWith}\rangle$ \\
            \end{tabular} & \{\textsf{celf}\} \\
    \end{tabular}
    }
    \label{tab:missed_path}
\end{table}

So far we have seen that the predicate path model presented in this work is able to accurately and quicly check the validity of statements of fact. Perhaps the most important contribution of this work, is not just in the ability to check facts, but rather in the ability to explain the meaning of some relationship between entities. Current progress in knowledge and reasoning in artificial intelligence is limited by our inability to understand the meaning behind data. For instance, although neural network-based technologies, like TransE, can produce accurate results, their learning mechanism does not provide an easily interpretable explanation for their answers. In contrast, our model explicitly provides a commonsense reason as to why a fact is deemed to be true or false. Table~\ref{tab:missed_path} shows some of the top predicate paths that are found by our model; we argue that they are generally intuitive and describe at least one key property about the given statement of fact.

One particularly interesting finding from Table~\ref{tab:missed_path} is the predicate path: \{\textsf{vice president}\} $\langle$\textsf{successor}, $\textsf{president}^{-1}\rangle$ \{\textsf{president}\}, which encodes, for example, that eventual-President Andrew Johnson succeeded Hannibal Hamlin as the second vice president under President Abraham Lincoln. Indeed, 8 US presidents have had two or more vice presidents (one succeeding the other) that have gone on to become president, meaning that the US constitution allows for the possibility to replace one vice president with another -- a little known, yet valid part of the definition of what it means to be the US vice-president.



\section{Related Work} \label{sec:related_work}



\vspace{5pt}\noindent\textbf{Discriminative Path Generation}. Although meta paths have been used in many methods such as similarity search, clustering, semi-supervised learning, and link prediction~\cite{Shi2014,Sun2011,Sun2012,zhao2015automatic,Lao2010}, these algorithms either require human annotated meta paths~\cite{fu2016graph} or enumerate all possible meta paths in the graph. Recently efforts have been made to meta path discovery and association rule mining in concept graphs~\cite{ruiz2014generating}, but most of the approaches have their own limitations. Meng~\etal, proposed a meta path generation algorithm that prunes the enumeration space by logistic regression, but this approach is prone to premature rejection and may miss important discriminative paths~\cite{Meng2015}. AMIE~\cite{Galarraga2013} is a global association rule mining algorithm which can not mine personalized,~\ie, context dependent, association rules as we shown in Sec.~\ref{sec:introduction} and \ref{sec:experiment}. Abedjan and Naumann proposed a predicate expansion algorithm~\cite{Abedjan2013} which can find predicate synonyms, but cannot find predicate paths that have discriminating power. The proposed discriminative path discovery framework in this work extracts meta paths and predicate paths from the graph directly with given endpoints, therefore our framework will not miss important predicate paths in the graph and is context-sensitive.

\vspace{5pt}\noindent\textbf{Fact Checking}. With the large volume of data generated every day, the number of unverified statements begets the need for automated fact checking~\cite{Graves2012,Cohen2011}. To that end, many researchers have focused on automated fact checking in recent years. Finn~\etal~introduced a new interactive tool to help human fact checkers determine the quality of a statement by extracting the propagation of facts on Twitter~\cite{Finn2014}. Ennals~\etal~created a crowd-sourced platform that highlight disputed claims~\cite{Ennals2010}. Kwok~\etal~proposed a ensemble method utilizing the result from search engines to check a given statement~\cite{Kwok2001}. Hassan~\etal~proposed a numerical fact monitor, called FactWatcher~\cite{Hassan2014}, that uses an append only database and certain skyline operators~\cite{Wu2012,Jiang2011}, but FactWatcher is not applicable to knowledge graphs or nonnumerical statements. The True Knowledge System~\cite{TunstallPedoe2010} validates a statement of the fact using $1,500$ predefined and user provided association rules; unfortunately, this means that it is impossible to check a statement that does not already have a predefined association rule within True Knowledge. Ciampaglia~\etal~published a knowledge graph based fact checking algorithm~\cite{Ciampaglia2015} utilizing node connectivity, but does not take advantage of the type-labels in the heterogeneous information networks. Recently, Guu~\etal~published a question answering algorithm that converts a given question into a vector space model to find the answer~\cite{Guu2015}, but, like neural network based models~\cite{Mikolov2013}, the learned model is generally uninterpretable. Li~\etal~proposed T-verifier, a search engine based fact checker~\cite{Li2011}, but such approach needs extensive access to search engine APIs which is not easy to gain. Knowledge graph completion methods, such as TransE~\cite{Bordes2013}, TransR~\cite{Lin2015}, and NTN~\cite{Socher2013} is not ideal for fact checking in large knowledge graphs because the slow convergence rate.

\vspace{5pt}\noindent\textbf{Link Prediction}. Apart from classic homogeneous link prediction methods, such as Adamic/Adar~\cite{Adamic2003}, SimRank~\cite{Jeh2002}, Katz~\cite{Katz1953}, Preferential Attachment~\cite{Barabasi1999}, and Personalized PageRank~\cite{Haveliwala2002} {\em etc.}, many heterogeneous methods have been developed to leverage the rich information in heterogeneous information networks. Heterogeneous graphlet base methods~\cite{Lichtenwalter2012} predict the relation between two endpoints by counting the occurrence of certain heterogeneous motifs, which are not applicable to complex knowledge graphs due to the exponential number of possible graph motifs. Other heterogeneous link prediction methods that adapt from classic homogeneous algorithms, HeteSim~\cite{Shi2014} and PCRW~\cite{Lao2010}, depend on human annotated meta paths. PathSim~\cite{Sun2011}, a heterogeneous similarity metric, also requires hand crafted and symmetric meta paths as the input. Recently Dong~\etal, proposed a heterogeneous link prediction algorithm based on coupled networks, but also needs human annotated meta paths as input~\cite{Dong2015}. In contrast, this work automatically discovers the important meta paths and predicate paths that related to the given statement of fact.

\section{Conclusions and Future Work}
\label{sec:conclusions}

We presented a fact checking framework for knowledge graphs that discovers the definition of a given statement of fact by extracting discriminative predicate paths from the knowledge graph, and uses the discovered model to validate the truthfulness of the given statement.

To evaluate the proposed method, we checked the veracity of several thousand statements across 8 different tasks on DBpedia and SemMedDB. We found that our framework was the all around best in terms of fact-checking performance and has a running time similar to existing models. We further tested the robustness of our algorithm by examining different ratios of true to false information and found that our framework was generally invariant to the class ratio. Finally, we showed that the proposed framework can discover interpretable and informative discriminative paths that are missed by other methods.

As this framework is the first of its kind, we leave much as a matter for future work. The next steps that are immediately obvious to us include extensions to this framework that perform (1) predicate identification, (2) enhanced entity representation, and (3) fact qualification. Predicate identification is the most natural extension to this framework wherein unnamed and unknown relationships can be implied through transitivity; for example, if a set of highly discriminative predicate paths between two sets of entities $x$ and $y$ exists, along with another set of highly discriminative predicate paths between $y$ and a third set of entities $z$, then we may be able to encode some special, transitive relationship between the entities in $x$ with the respective entities in $z$. Because we are encoding the meaning behind relationships and between entities, it is likely that we will be able to find natural implications that arise in arithmetic combinations of entities such as the canonical \textsf{King}-\textsf{man}+\textsf{women}=\textsf{Queen}, but with a human-interpretable representation for each operator that is not present in current vector-based models. Finally, we should be able to use mismatches and errors in our model to qualify some statement of fact; for example, the statement ``Rome is the capital of the Roman Empire'' is only true before 323 CE, after which the capital was changed to Constantinople.

\section{Acknowledgments}
This work is sponsored by an AFOSR grant FA9550-15-1-0003, and a John Templeton Foundation grant FP053369-M.

\section{References}
%

\end{document}